\documentclass[lettersize,journal]{IEEEtran}
\usepackage{amsmath,amsfonts}
\usepackage{algorithm}
\usepackage[noend]{algpseudocode}
\usepackage{array}
\usepackage[caption=false,font=normalsize,labelfont=sf,textfont=sf]{subfig}
\usepackage{textcomp}
\usepackage{stfloats}
\usepackage{url}
\usepackage{verbatim}
\usepackage{graphicx}
\usepackage{cite}

\usepackage[numbers,sort&compress,square]{natbib}

\usepackage{graphics}
\usepackage{epsfig}
\usepackage{times}
\usepackage{caption}
\usepackage{array}
\usepackage{booktabs}
\usepackage{multirow}
\usepackage{mathtools, nccmath}
\usepackage{color}

\usepackage{amsmath}
\usepackage{amssymb} 
\usepackage{svg}
\usepackage{makecell}

\begin{document}

\title{Decision Making in Urban Traffic: A Game Theoretic Approach for Autonomous Vehicles Adhering to Traffic Rules}

\author{Keqi Shu,~\IEEEmembership{Graduate Student Member,~IEEE,}  Minghao Ning$^*$, Ahmad Alghooneh,  Shen Li, Mohammad~Pirani,~\IEEEmembership{Member,~IEEE,} Amir~Khajepour,~\IEEEmembership{Senior Member,~IEEE,} 
        % <-this % stops a space
\thanks{The authors would like to acknowledge the financial support of National Sciences and Engineering Research Council of Canada (NSERC) in this work.  

$*$ corresponding author, email: minghao.ning@uwaterloo.ca}% <-this % stops a space
%\thanks{Manuscript received April 19, 2021; revised August 16, 2021.}

}

% The paper headers
\markboth{IEEE TRANSACTIONS ON INTELLIGENT TRANSPORTATION SYSTEMS}%
{Shell \MakeLowercase{\textit{et al.}}: A Sample Article Using IEEEtran.cls for IEEE Journals}

%\IEEEpubid{0000--0000/00\$00.00~\copyright~2021 IEEE}
% Remember, if you use this you must call \IEEEpubidadjcol in the second
% column for its text to clear the IEEEpubid mark.

\maketitle

\begin{abstract}
One of the primary challenges in urban autonomous vehicle decision-making and planning lies in effectively managing intricate interactions with diverse traffic participants characterized by unpredictable movement patterns. Additionally, interpreting and adhering to traffic regulations within rapidly evolving traffic scenarios pose significant hurdles. This paper proposed a rule-based autonomous vehicle decision-making and planning framework which extracts right-of-way from traffic rules to generate behavioural parameters, integrating them to effectively adhere to and navigate through traffic regulations. The framework considers the strong interaction between traffic participants mathematically by formulating the decision-making and planning problem into a differential game. By finding the Nash equilibrium of the problem, the autonomous vehicle is able to find optimal decisions. The proposed framework was tested under simulation as well as full-size vehicle platform, the results show that the ego vehicle is able to safely interact with surrounding traffic participants while adhering to traffic rules.
\end{abstract}

\begin{IEEEkeywords}
Autonomous vehicles, intersection handling, decision making, game theory, traffic rule, real-world vehicle.
\end{IEEEkeywords}

\section{Introduction}
\subsection{Background}
\IEEEPARstart{A}{utonomous} vehicles (AV) are getting more and more involved into our daily lives,  fundamentally altering how people navigate on the road. With the advancement of self-driving technologies, AVs are gradually transitioning to SAE-standard level 2 and even level 4 autonomy. However, as these autonomous vehicles brings convenience to our daily lives, there still lies a number of issues which causes trepidation of risks of unsafe actions or weird behaviours under urban driving. 

One significant issue revolves around the interaction between humans and AVs. In urban driving scenarios, vehicles, cyclists, and pedestrians often operate in close proximity, with each entity's behaviour significantly influencing other. Addressing this complex interaction not only consumes time but also demands accurate modelling, posing a formidable challenge.

Another challenge arising from urban driving pertains to the adherence of AVs to traffic rules. These rules are crucial for ensuring the safety of both the AV itself and other road users. However, interpreting these rules into programming language poses significant challenges due to their complexity. Furthermore, the dynamic nature of urban environments means that right-of-way situations can change rapidly, further complicating matters.

These challenges, along with others, contribute to the complexity of self-driving in urban scenarios. Overcoming these obstacles is essential for achieving full autonomy in autonomous vehicles.

\subsection{Literature review}
Over the past decade, researchers have developed numerous algorithms and frameworks aimed at tackling the challenges present in highly interactive urban driving scenarios \cite{shi2023trajectory, gu2015tunable}. Crosato $et\ al.$ considered the interaction among traffic participants using social value orientation and deep reinforcement learning, which improves the accuracy in modeling human behaviors under interactive scenarios \cite{crosato2022interaction}. Wang $et\ al.$ introduced a latent space reinforcement learning method to address the strong interaction among traffic participants. The application of hidden Markov models and Gaussian mixture regression allows vehicles to find a balance between driving safety and efficiency \cite{wang2021interpretable}. Ding $et\ al.$ proposed a hierarchical decision-making model under strong interaction with a partially observable Markov decision process, then they applied a spatio-temporal semantic corridor model for motion planning, which allows the vehicle to avoid over-conservative motions \cite{ding2021epsilon}.

These methods have shown promising results in complex urban driving scenarios; however, in real-world applications, the aforementioned algorithms are less efficient at considering the two-way interaction of traffic participants in a mathematical way. To formulate the interaction more realistically, the ego vehicle not only needs to mathematically consider the influence from surrounding vehicles on itself but also must consider its own behavior's impact on surrounding vehicles.
In highly interactive scenarios, game theory emerges as a valuable tool for mathematically formulating this two-way interaction behaviours. By finding the Nash equilibrium, the ego vehicle can determine an optimal solution to navigate through traffic \cite{yuan2024game}.

Hang $et\ al.$ formulated the interaction between AVs and human drivers using a non-cooperative Stackelberg game, combined with model predictive control to generate predicted behaviours for the AV \cite{hang2020human}. Zanardi $et\ al.$ modeled the interaction between traffic participants with an Urban Driving Games model, enabling the AV to find socially efficient equilibrium under an iterated best response scheme \cite{zanardi2021urban}. \cite{vicini2022decision} considered strong interaction under emergency scenarios, where two vehicles plan safe paths while avoiding randomly moving obstacles using game theory \cite{vicini2022decision}.

Shu $et\ al.$ proposed a differential game to handle AVs' intersection decision-making, considering various types of traffic participants and extracting different features from real-world data to better estimate the types of traffic participants \cite{shu2023human}. \cite{zhang2023game} also incorporated the social characteristics of traffic participants using a Divergent Social Value Orientation model and solved interactive lane-change decision-making with a game theory model. Ji $et\ al.$ discovered that a social gap exists for strongly interactive lane-change behaviours by formulating the lane-change decision-making as a game \cite{ji2020estimating}.

Jeong considered the uncertainties surrounding drivers' unknown intentions during interactions with human drivers at uncontrolled intersections. The author applied a Multiple Model filter to estimate the uncertain intentions of human drivers, which were then used in the game-theory-based decision maker to generate appropriate decisions for the ego vehicle \cite{jeong2023probabilistic}. Rahmati $et\ al.$ also accounted for uncertainties in the environment and drivers' behaviour at intersections when formulating the decision-making problem as a game. The proposed framework has been shown to improve AV behaviour at intersections \cite{rahmati2021helping}.

Game theory is a powerful tool to help AVs consider the strong interactions between human traffic participants and AVs while making decision. However, in the real world, strong interaction between traffic participants needs to be considered along with obeying proper traffic rules in order to assure safety while achieving optima decision. 

Incorporating traffic rules into decision-making models is complicated due to their difficulty in being formulated into mathematical logic operating within well-defined parameters. This complexity of interpreting traffic rules arises from the dynamic nature of traffic scenarios and the fast-changing right-of-way accordingly. Thus, merging traffic rules into mathematical game theory frameworks is challenging.

Liu $et\ al.$  considered soft traffic rules, such as first-come-first-go, in their decision-making framework and used game theory to decide whether AVs should violate soft rules in emergency scenarios \cite{liu2022three}. Shu $et\ al.$  implemented traffic rules in their game-theory-based decision-making framework by considering speed limits for each traffic participant \cite{shu2023human}.

Fan $et\ al.$ studied the interaction between various traffic participants at uncontrolled intersections, including both traffic obeyers and violators, to improve the traffic efficiency of vehicles at intersections \cite{fan2022food}. \cite{bjornskau2017zebra} compared the differences between game-theoretic modeling and strict traffic rule modification. The results showed that game theory could be a useful tool for better understanding traffic rules.

\begin{figure*}
    \centering
    \includegraphics[width=0.8 \textwidth]{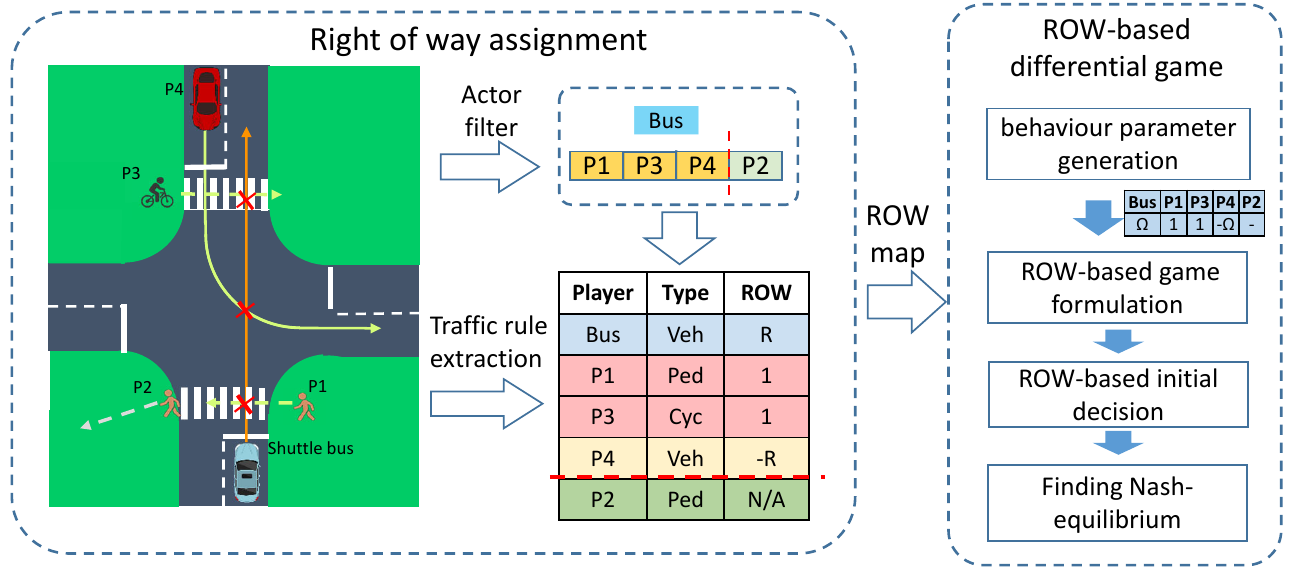}
    \caption{Naturalistic autonomous intersection handling framework}
    \label{Fig: framework}
\end{figure*}

The literature mostly focuses on specific traffic rules within constrained scenarios, such as adhering to a first-come-first-go sequence at intersections or obeying speed limits. However, the diversity of traffic rules across various scenarios presents significant challenges. This complexity necessitates models that capture not only the nuances of local regulations but also the dynamic interactions between vehicles operating under different rules. Therefore, developing a more generalized model for traffic rule extraction and mathematical representation could significantly enhance the implementation of traffic rules in typical urban driving scenarios.
%The aforementioned literature consider a specific traffic rule in a certain scenario or simple common traffic rule in a constrained scenario, such as following the first-come-first-go sequence at an intersection or driving under speed limit. However, rules at various traffic scenarios could be very diversed, this diversity requires models that not only incorporate the nuances of local regulations as well as the dynamic interactions between vehicles adhering to different rules, which is very challenging. 
%therefore, a more generalized traffic rule extraction model would improve the implementation of traffic rules in common urban driving scenarios. The ability of mathematically consider complex traffic rules in strong-interactive scenarios still remain a challenge. 
%Consideration of rules in the game-theory-based decision-making framework is still in an incipient stage, such as applying static rules like speed limits into the game, or considering single rules such as first-come-first-go. Incorporating more complex and comprehensive rules into the game-theory framework remains a yet-to-be-solved issue.

\subsection{Contribution}
This paper addresses the issue of autonomous vehicles' decision-making processes within highly interactive urban driving scenarios featuring complex traffic rules. It proposes a rule-adherent, game-based decision-making framework. This framework dynamically accommodates complex rules across various urban traffic scenarios, including all-way stop sign intersections, T-junctions, and roads with pedestrian crosswalks. It extracts right-of-way based on road geometry and the states of the traffic participants, then converts this information into behavioral parameters. This allows the game formulation of the strong interaction considering traffic rules. By solving the formulated differential game, the ego vehicle can identify optimal decisions that balance safety and commuting efficiency while adhering to general traffic rules.
%It is designed to generate right-of-way assignments for each traffic participant according to traffic rules and then assigns corresponding behavioural parameters. These parameters are utilized to formulate a differential game between the ego vehicle and surrounding traffic participants, enabling the ego vehicle to account for strong interactions in urban scenarios. By solving this game, the ego vehicle can identify optimal decisions that prioritize safety and commuting efficiency while adhering to general traffic rules.
The contributions of this paper are:
\begin{enumerate}
    \item Proposed a game-theory-based decision-making framework that dynamically considers various kinds of traffic rules among participants in different strong-interaction scenarios. This results in a more realistic interaction formulation, aimed at achieving more human-like decision-making in urban areas.
    \item Proposed a generalized traffic rule interpretation model that mathematically translates rules into behavior parameters using their right-of-way as the median. This allows the proposed framework to dynamically incorporate traffic rules in various kinds of scenarios.
    \item Conducted verification of the rule-adherent, game-theory-based decision-making model using a real-world shuttle bus operating in urban areas.
\end{enumerate}

\section{Framework overview}
The proposed framework utilizes the right-of-way of various traffic participants as a means to incorporate traffic rules into the decision-making process of autonomous vehicles. Fig.~\ref{Fig: framework} illustrates the framework of rule-based decision-making for autonomous vehicles.

On the left side of Fig.~\ref{Fig: framework}, the process by which the ego vehicle extracts environmental information and identifies the locations of surrounding traffic participants to generate lists of right-of-ways for participant is depicted. The ego vehicle initially applies an actor filter to all traffic participants, enabling it to prioritize those with which it interacts with. This filter takes into account the future paths of each traffic participant and identifies them as critical if their future paths overlap with the ego vehicle's projected path based on time. In the scenario depicted in the upper-left corner of the left-side box of Fig.~\ref{Fig: framework}, the blue vehicle represents the ego autonomous vehicle traveling straight, while surrounding traffic participants are also depicted along with their future paths.

In the top right part of the left-side box, the filtered results are displayed. The ego vehicle is represented by a blue rectangle labeled ``Bus". Among the identified traffic participants, the oncoming vehicle (P4), cyclist (P3), and one pedestrian (P1) have future trajectories overlapping with the ego vehicle, making them critical participants, and are therefore labeled as yellow rectangles. Conversely, the pedestrian (P2) walking away from the road is deemed less critical and is labeled as a green rectangle.

\begin{figure}
    \centering
    \includegraphics[width=0.36 \textwidth]{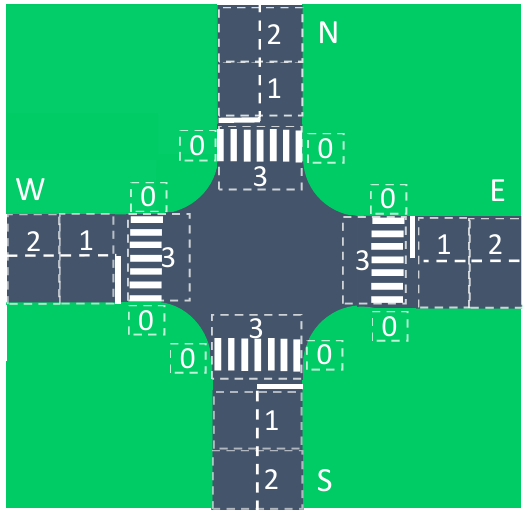}
    \caption{Traffic participants' minimum distances distribution}
    \label{Fig: ROWScenario}
\end{figure}

Following the filtering process, right-of-way is assigned to each participant by considering their current position and actor type according to traffic rules. For instance, pedestrians and cyclists are granted absolute priority, while vehicles traveling straight are given higher priority compared to those turning left. Adhering to these rules, right-of-way is assigned to each traffic participant and presented in the table located at the bottom right of the figure. 

The framework can incorporate the right-of-way of each traffic participant and consider the successive interactions between them, as illustrated in the right-side box of Fig.~\ref{Fig: framework}. By formulating the decision-making problem as a differential game, the framework addresses the high-interaction behaviour among traffic participants. The assigned right-of-way for each participant is translated into behaviour parameters, which influence the assumptions about the behaviour of surrounding traffic participants within the differential game setup.

The blue table on the right-side box of Fig.~\ref{Fig: framework} displays the result of the right-of-way table transferred into behaviour parameters. The traffic participants with absolute high right-of-way are assigned the highest value, while the behaviour parameter value for the remaining participants decreases proportionally with their right-of-way. Subsequently, the rule-based differential game is formulated, and by finding the Nash equilibrium of the differential game given an initial policy, the optimal decision of the ego vehicle can be calculated. A more detailed description of the proposed framework will be presented in the following sections.

\section{Traffic rule extraction}
Traffic rules are a fundamental component of safe and efficient driving for autonomous vehicles. However, interpreting traffic rules into decision-making frameworks can be challenging due to their ambiguity and rapidly changing nature in complex traffic scenarios. The proposed framework addresses this challenge by extracting right-of-ways from the positions and types of traffic participants, and translating these right-of-ways into behaviour parameters for the differential game.

In real-world driving scenarios, traffic rules can range from definite and clear instructions, such as yielding at red traffic lights or to pedestrians, to situations where rules provide indefinite precedence for traffic participants crossing the road, such as at intersections with yield signs or uncontrolled intersections. To effectively interpret both definite and indefinite precedence from traffic rules in urban driving, the extraction of right-of-way is essential to adapt to this complexity.

The extracted right-of-way includes absolute high or low, as well as relative high or low right-of-way, corresponding to the definite and indefinite precedence from traffic rules. Proper extraction of right-of-way for different traffic participants necessitates consideration of their type, location, and the specific traffic scenario.

For instance, pedestrians and cyclists on or near the road are assigned the highest right-of-way, while vehicles on the side of the road with a yield sign have lower right-of-way compared to those on the main road.

The primary process of assigning right-of-way is as follows: first, a list of interactive traffic participants are filtered using the actor filter. Then, the type of these traffic participants is considered, with pedestrians and cyclists having absolute high right-of-way. All other traffic participants are categorized as having relative high, low, or neutral right-of-way.
\begin{table}[]
\centering
\caption{Right-of-way and corresponding behavioral parameters of traffic participants in various scenarios} 
\begin{tabular}{cccccc}
                               & Ego                    & Veh1                  & Veh2             & Ped & Cyc \\ \Xhline{1pt} %\hline
\multirow{3}{*}{4-way stop sign}    & E1-S                   & N2-S                  & \textbackslash{} & E3  & S0  \\ \cline{2-6} 
                               & Neu                      & Abs L                    & \textbackslash{} & Abs H   & Neu   \\ \cline{2-6}
                                & 0                      & -1                    & \textbackslash{} & 1   & 0   \\ \Xhline{1pt}
\multirow{2}{*}{Uncontrolled}  & E1-S                   & W2-L                  & N2-R             & S3  & N3  \\ \cline{2-6} 
                                & H                & L                & L           & Abs H   & Abs H   \\ \cline{2-6} 
                               &  $0.05$                & $-0.1$                & $-0.1$           & 1   & 1   \\ \Xhline{1pt}
\multirow{2}{*}{WE stop sign}  & E1-S                   & N1-S                  & \textbackslash{} & N3  & S0  \\ \cline{2-6} 
                                 & Abs L                     & Neu                     & \textbackslash{} & Abs H   & Neu   \\  \cline{2-6} 
                               & -1                     & 0                     & \textbackslash{} & 1   & 0   \\ \Xhline{1pt}
\multirow{2}{*}{WE yield sign} & E1-S                   & N1-S                  & \textbackslash{} & N3  & S0  \\ \cline{2-6} 
                                & L & H & \textbackslash{} & Abs H   & Neu   \\ \cline{2-6} 
                               & $-0.05$ & $0.05$ & \textbackslash{} & 1   & 0   \\ \Xhline{1pt}
\end{tabular}
\label{Tab: ROW}
\end{table}
\begin{figure}
    \centering
    \includegraphics[width=0.5 \textwidth]{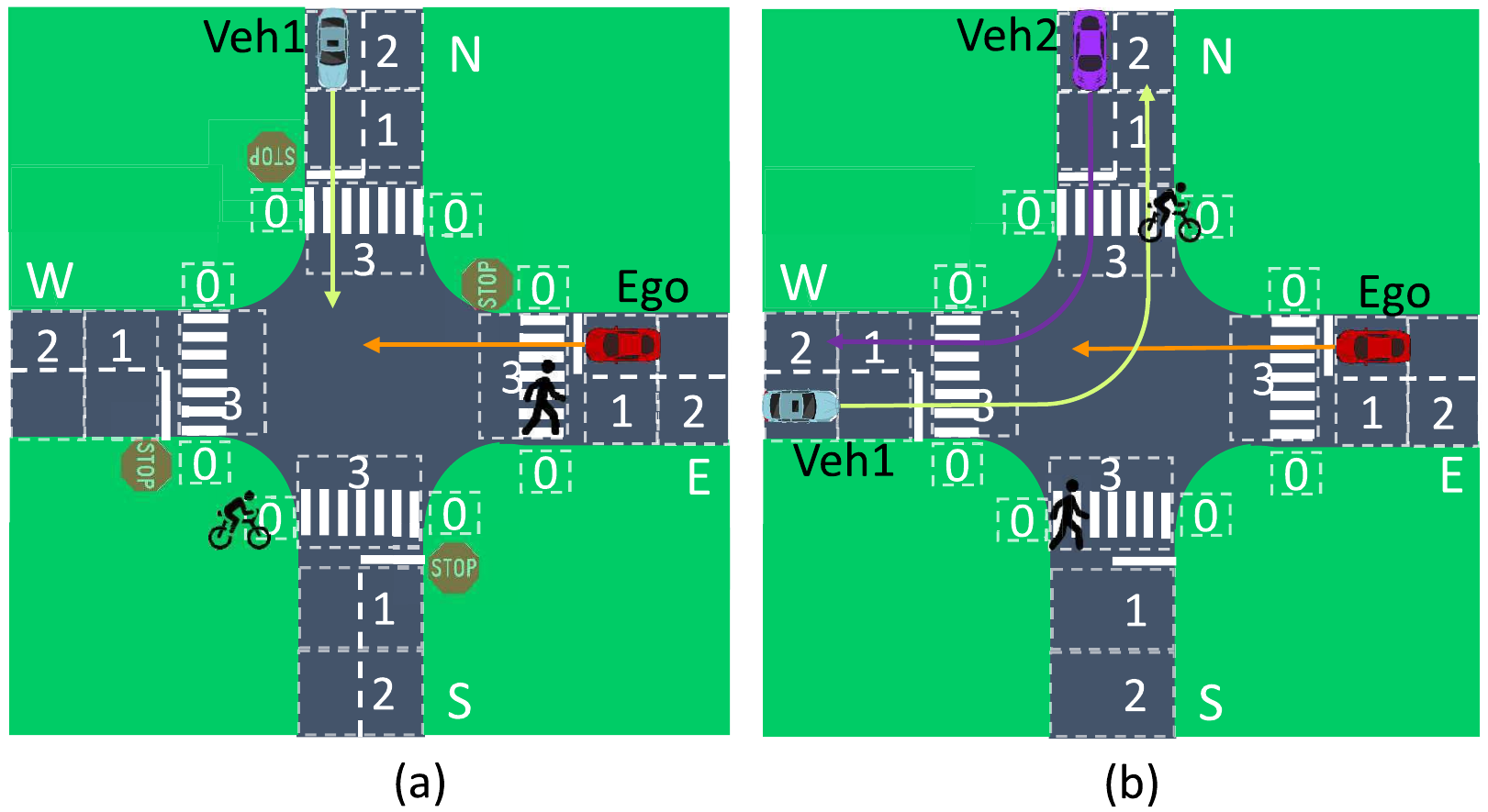}
    \caption{Right-of-way extraction from typical traffic scenarios}
    \label{Fig: ROWSceExample}
\end{figure}

Next, the traffic scenario is considered. For instance, at an uncontrolled intersection with no designated main road, pedestrians and cyclists are granted absolute high right-of-way. Vehicles, on the other hand, are assigned a relative right-of-way that is either high, low or neutral, based on their distance to the center of the intersection and their projected future paths. Conversely, in a controlled intersection with traffic lights, vehicles on lanes with a red light are assigned absolute low right-of-way. Among vehicles on lanes with a green light, those proceeding straight are given relative high right-of-way, while those making left or right turns are assigned relative lower right-of-way. %More possible traffic scenarios will be presented and evaluated in the rest of the section.

In an intersection with a yield sign, where one way is designated as the main road and the other as the secondary road, vehicles on the main road are assigned higher right-of-way than those on the secondary road. At an all-way stop sign intersection, right-of-way is assigned based on the distance of each vehicle to the stop line. The vehicle closest to the stop line is set as having neutral right-of-way, while the rest of the vehicles are assigned absolute low right-of-way. In cases where two vehicles arrive at an intersection simultaneously, according to Ontario, Canada's traffic regulations, the vehicle on the right has higher right-of-way. Therefore, in such instances, the right-side vehicle will have a relatively higher right-of-way than the vehicle stopped on the left.

To illustrate the extraction of right-of-way for different traffic participants, various possible traffic scenarios are presented in this section in Fig.~\ref{Fig: ROWScenario}, along with the extracted right-of-way shown in Table~\ref{Tab: ROW}. These scenarios are depicted using Fig.~\ref{Fig: ROWScenario}, where ``N", ``W", ``S", and ``E" denote the sides of the road at the intersection, and the dash-contour blocks labeled with numbers 0 to 3 represent the potential positions of each traffic participant.

The corresponding Table~\ref{Tab: ROW} describes the scenario, the first column indicates the traffic environment, e.g., all-way stop sign intersection and uncontrolled intersection. For each traffic environment, the first row indicates the position and predicted future path of each traffic participant, e.g., the Ego vehicle in the first 4-way stop sign environment is at the east side of the intersection at the block with number 1 on it (ones closer to the stop line.). The second and third rows represent the corresponding right-of-way and behavioural parameters, respectively. Subsequent columns indicate the other traffic participants, including the ego vehicle, human-driven vehicle 1, human-driven vehicle 2, pedestrian, and cyclist.

To better demonstrate the extraction of the right-of-way in various traffic scenarios, the first two scenarios listed in Table~\ref{Tab: ROW} are visually represented in Fig.~\ref{Fig: ROWSceExample}. Fig.~\ref{Fig: ROWSceExample} (a) illustrates the first traffic scenario, and Fig.~\ref{Fig: ROWSceExample} (b) corresponds to the second scenario from Table~\ref{Tab: ROW}.

In the first scenario, a 4-way stop-sign intersection described in Table~\ref{Tab: ROW}, the pedestrian on the east side crosswalk of the intersection is assigned absolute high right-of-way. The cyclist, not within the intersection’s drivable space, is designated with neutral right-of-way. Vehicle 1 (blue vehicle in Fig.~\ref{Fig: ROWSceExample} (a)), being farther from the stop line compared to the ego vehicle (red vehicle in Fig.~\ref{Fig: ROWSceExample} (a)), is assigned absolute low right-of-way, while the ego vehicle's right-of-way is set as neutral.

In the second scenario, an uncontrolled intersection, the right-of-ways for the vehicles are determined by their proximity to the stop line. Vehicles 1 (blue vehicle in Fig.~\ref{Fig: ROWSceExample} (b)) and 2 (purple vehicle in Fig.~\ref{Fig: ROWSceExample} (b)), positioned in section 2 and farther from the stop line, are assigned relative low right-of-way. The ego vehicle (red vehicle in Fig.~\ref{Fig: ROWSceExample} (b)), closer to the stop line in section 1, is granted relative high right-of-way. Both the cyclist and pedestrian on the crosswalk receive absolute high right-of-way.

For the remaining two traffic environments presented in Table~\ref{Tab: ROW}, which include scenarios with stop signs and yield signs on the WE-side of the road, similar methods can be employed to assign right-of-way.

After extracting the right-of-way of each traffic participant, they are then converted into behavioural parameters that the game-theory-based decision-making framework can utilize to generate behaviours while considering traffic rules. behavioural parameters, denoted as $\gamma$, range from -1 to 1, where 1 corresponds to traffic participants with absolute high right-of-way, and -1 corresponds to absolute low right-of-way.

For traffic participants with relative high or low right-of-way, their behavioural parameters are mapped between -0.25 and 0.25. In general intersection scenarios, behavioural parameters are assigned based on the distance from the vehicle to the stop line in a discrete manner with a step of 0.05. The behavioural parameters for vehicles with relative high right-of-way are positive and increase as they approach the intersection, while those for vehicles with relative low right-of-way are negative and increase as they approach the intersection.

\begin{equation}\label{equ:ROWtoBP} 
        \gamma=\begin{cases} \ \ \frac{1}{20} (N_s - \lceil \frac{d_{dist}}{d_{len}}  \rceil +1), & \text { if }  ROW = H \\
        -\frac{1}{20}(N_s - \lceil \frac{d_{dist}}{d_{len}}\rceil+1) , & \text { if } ROW = L\\\  \ 1 , & \text { if }  ROW = Abs\  H
        \\-1 , & \text { if }  ROW = Abs\ L
        \\\ \ 0 , & \text { if }  ROW = Neu
        \end{cases},
\end{equation}

(\ref{equ:ROWtoBP}) demonstrates how the right-of-way is transferred into behaviour parameters. $\gamma$ indicates the behavior parameter, $ROW$ indicates the right-of-way, $d_{dist}$ indicates the distance from the vehicle to the intersection stop line. $d_{len}$ indicates the length of each road section presented in Fig.~\ref{Fig: ROWScenario} (e.g., length of part "1"), and $N_s$ is the number of road sections on each side of the road, which is always smaller than 6. For example, in Fig.~\ref{Fig: ROWScenario}'s scenario, $N_s = 2$. For each vehicle, the behavior parameter is calculated by rounding up the value of $\frac{d_{dist}}{d_{len}}$ if the right-of-way is relatively high or low ($ROW = H/L$).

Through (\ref{equ:ROWtoBP}), behaviour parameters can be computed and are presented in the third row of each scenario in Table~\ref{Tab: ROW}. A more detailed explanation of how the behaviour parameters influence the decision-making framework will be provided in the next section.

\section{Rule-based differential game}
In order to consider interaction behaviours in a mathematical way between traffic participants, the framework formulates the decision making problem into a game. This game-theory-based framework assesses the potential behaviours of surrounding traffic participants and evaluates the impact of the ego vehicle's decisions on these participants, enabling it to make informed decisions.

Moreover, to account for traffic rules, the behaviour parameters derived from the right-of-way are integrated into the game. This integration enables the ego vehicle to account for the various interactions and behaviours of traffic participants dictated by the traffic rules. By iteratively seeking the optimal solution of the game, a Nash equilibrium can be reached, leading to the optimal decision for the ego vehicle.

\subsection{Game formulation}

The dynamics of the traffic participants must be accurately defined to generate precise estimates of path and speed. The dynamic models utilized in this framework encompass both 5-dimensional and 4-dimensional models, expressed in a general format as $\dot{s}(t) = f\left(s(t), u_{1:N}(t)\right)$, where $s$ represents the state space and $u$ denotes the action space.

The 5-dimension model is used to formulate the dynamic for vehicles and cyclists could be written as:
\begin{equation}\label{chp:vehicle_dynamic}
\dot{s}(t) {=}\setlength\arraycolsep{1pt}
\begin{bmatrix} \dot{x} \\ \dot{y} \\ \dot{\theta} \\ \dot{\phi} \\ \dot{v} \end{bmatrix} {=} \underbrace{\begin{bmatrix} 0 & 0 & -v\sin{\theta} & 0 & \cos{\theta} \\ 0 & 0 & v\cos{\theta} & 0 & \sin{\theta} \\ 0 & 0 & 0 & \frac{v\tan{\phi}}{l\phi}  & 0 \\ 0 & 0 & 0 & 0 & 0 \\ 0 & 0 & 0 & 0 & 0 \end{bmatrix}}_A \begin{bmatrix} x \\ y \\ \theta \\ \phi \\ v \end{bmatrix} + \underbrace{\begin{bmatrix} 0 & 0 \\ 0 & 0 \\ 0 & 0 \\ 1 & 0 \\ 0 & 1 \end{bmatrix}}_B \begin{bmatrix} \dot{\phi} \\ \dot{v} \end{bmatrix}
\end{equation}
where the state space comprises $(x, y, \theta, \phi, v)$,  representing the position in the $x$ and $y$ coordinates, yaw, steering angle, and speed of the traffic participant, respectively. The action space is denoted as $[a_1, a_2]^T = [\dot{\theta} , \dot{v}]^T$, representing the steering rate and acceleration, respectively. Finally, matrices A and B represent the state and action transfer matrices. The path and speed generated using this model are continuous, and the curvature of the generated path also conforms to the length of the vehicle or cyclist.

The dynamic model for pedestrians is a 4-dimension model similar to (\ref{chp:vehicle_dynamic}), only the states dose not consider the steering angle, and the actions space includes the acceleration and yaw rate of the pedestrian. Both the 4-dimensional and 5-dimensional models operate under continuous time and continuous space, which entails significant computational effort when considering the potential paths of traffic participants, as well as in finding the optimal path for the ego vehicle.

Henceforth, the dynamics of traffic participants are approximated using Jacobian linearization (Taylor expansion) \cite{grognard1999global}. This technique transforms the state space $s$ and action space $a$ of the original problem into variations of states and actions from their current values, denoted as $\delta s$ and $\delta u$ respectively:
\begin{equation}\label{equ: general_dynamic}
    \dot{\delta}_{s}(t)=A'(t) \delta{s}(t) + \sum_{i \in[N]} B_{i}'(t) \delta{u}(t),
\end{equation}
In this context, $A'$ and $B'$ represent the derivatives of the system dynamics with respect to the state and input, respectively. It's essential to note that when applying Taylor expansion, linearization is valid only if $\delta{s}(t)$ and $\delta{u}(t)$ are sufficiently small. This linearization also holds for discrete systems, where it retains the format of (\ref{equ: general_dynamic}), with $A'$ and $B'$ matrices replaced by $e^{\mathbf{A} T}$ and $\mathbf{A}^{-1}\left(e^{\mathbf{A} T}-\mathbf{I}\right) \mathbf{B}$ respectively. Leveraging linearized dynamics enables the generation of paths and speed profiles for traffic participants in a more efficient manner.

The framework models the characteristics and behaviours of all traffic participants by defining their utility functions as $g_{i} \left(t, s(t), u_{1:N}(t)\right)$, where $i$ represents the index of the traffic participant, $N$ is the total number of traffic participants, and $t$ denotes the time. With this utility function, the framework can assess the potential alternative paths and speed profiles from the current states.

The general utility function for all traffic participants is considered in four aspects: safety, commuting efficiency, comfort riding, and adherence to traffic rules. Thus, the utility function could be expressed as:
\begin{equation}\label{equ:reward_equ}
g_{i} \left(t, x(t), u_{1: N}(t)\right) = \sum_{i = 0}^{N} {[ K_{s}, K_{e}, K_{c}, K_{r}] _ i \left[\begin{array}{l}
R_{s} \\
R_{e} \\
R_{c} \\
R_{r}
\end{array}\right] _i},
\end{equation}
where $R_{s}$, $R_{e}$, $R_{c}$ and $R_{r}$ represent the cost for safety, commuting efficient, comfort riding and obeying traffic rule feature of the utility function, and the corresponding $K_{[\dot\ ]}$s are the relevant linear constant ratios. 

The utility functions of the traffic participants, as described in (\ref{equ:reward_equ}), characterize their behavioral tendencies using various features: $R_{s}$, $R_{e}$, $R_{c}$, and $R_{r}$. For instance, with the safety feature $R_{s}$, if traffic participants are closer than their minimum safety distance, a significant cost is incurred, indicating a tendency towards collision avoidance.
Therefore, the accuracy of behavioral modeling critically depends on the parameters and thresholds of these features such as minimum safety distance among traffic participants in $R_{s}$ and nominal speed in $R_{e}$ is critical in the accuracy of the behavioral modeling. To select proper parameters and thresholds, a real-world behavioral dataset was studied and utilized to derive the necessary values of these parameters and thresholds in our previous work \cite{shu2023human}. For a more detailed demonstration of the utilities for traffic participants, please refer to \cite{shu2023human}.

The traffic rule feature $R_{r}$ is set for vehicles (with absolute right-of-way) to follow traffic rule and fully stop before the stop sign or red light. This feature could be expressed as:
\begin{equation}\label{equ:reward_equ_Rule}
    R_{r}= \begin{cases} 0, 
    \\\quad\quad\quad \text { if } (\sum_{t=2}^T ||p_i'(t), p_i'(t-1 )||_2 
     -d_{ss}) \geq 0  
    %\\(p_i\in l_{center}) 
     \\ |\sum_{t=1}^T ||p_i'(t), p_i'(t-1 )||_2 - d_{ss}|^2,
    \\ \quad\quad\quad\text { if }  ( \sum_{t=2}^T ||p_i'(t), p_i'(t-1 )||_2 -d_{ss}) < 0\end{cases}
\end{equation}
%\begin{equation}\label{equ:reward_equ_Rule}
%    R_{l}= \begin{cases} 0, \\\quad\quad\quad\quad \text { if } dis(p_{i}%(x,y), l_{poly}) < l_{b} \\ {k_l}(\min_{t=0}^T proj (p_{i}(x(t),y(t)),l_{poly})- l_{b})^2 ,  \\\quad\quad\quad\quad\text { if }  dis(p_{i}(x,y),l_{poly}) \geq l_{b}\end{cases},
%\end{equation}
where $p'_i(t)$ is the closet point from the vehicle's position $p_i(t)$ to the center line of the drivable space $l_{poly}$,  and $i$ is the index of vehicle in the scenario. While $t\in (0,T)$ represents the corresponding time of that position, and $T$ is the planning horizon. For a vehicle approaching the stop sign, $l_{poly}$ would correspond to the center line of the road from the current position of the ego vehicle to the stop line of the intersection. $d_{ss}$ is the length of  $l_{poly}$ and $||\cdot||_2$ presents the $l-2$ norm between $p'_i(t)$ and $p'_i(t-1)$. Please note that this feature applies only to the traffic participant that needs to stop in front of the stop line.
%where the $l_{poly}$ represents the center line of the drivable space. For a vehicle approaching the stop sign, this would correspond to the center line of the road from the current position to the stop line of the intersection. $p_{i}(x,y)$ denotes the position of the traffic participant, and $l_b$ is the half width of the road. $proj (p_{i}(x(t),y(t)),l_{poly})$ indicates the closest distance from the current planned path $p_{i}(x(t),y(t))$ of the $i$th vehicle along the planning horizon $t\in (0,T)$ to the center line of the drivable space $l_{poly}$. Finally, $k_l$ represents a large ratio for the utility. Please note that this feature applies only to the traffic participant that needs to stop in front of the stop line.

With this utility function, if any part of the planned path deviates from the drivable space, a large utility penalty will be assigned to the traffic participant. It's important to note that for traffic participants with constrained drivable space, the nominal path will be aligned with the center line of the drivable space for all other utility features.

For the safety feature, the primary focus is on maintaining adherence to road boundaries, operating within the speed limit, and maintaining safe distances from other traffic participants. These safe distances vary depending on the type of traffic participant and their behavioural style. The utility function will impose significant penalties on traffic participants not adhering to safe behaviour.

The commute efficient feature is calculated by applying mild penalties to speed profiles deviating from the nominal speed and to paths deviating from the current path. The aim is to encourage operation at the desired optimal speed and path, thereby promoting efficiency in commuting.

The comfort riding feature imposes subtle penalties on any acceleration or steering rate (yaw rate for pedestrians) applied to the traffic participant. This aspect aims to encourage smoother paths, %with fewer abrupt movements, 
enhancing comfort for all road users. For a more detailed expression and mathematical explanation of the safety, commute efficient, and comfort riding utility functions for these features, please refer to \cite{shu2023human}.

Given the utility functions of each traffic participant, the potential deviations from their nominal behaviours can be evaluated. To further account for interaction behaviours between traffic participants, the utilities are formulated in a linear and quadratic form, akin to the iterative linear quadratic game format \cite{fridovich2020efficient}. To incorporate traffic rules into the game, the right-of-ways are translated into behavioural parameters and applied to the game, allowing for the incorporation of definite and indefinite precedence as dictated by traffic rules.

The behavioural parameters are integrated into the linearized and quadraticized utility function to enforce rules:
\begin{equation}\label{equ:reward}
\begin{aligned}
    g(s+&\delta  s,{u}_{1: N}(k) + \delta u_{1:N}(k),k)  -  g(s,{u}_{1: N}(k),k) \approx \\& \frac{1}{2} h_i^{-1}(\gamma) \cdot\delta s(k)  \cdot Q_{i}(t) \cdot \delta s(k)^{T} + h_i^{-1}(\gamma) \cdot \delta s(k)^{T} \cdot l_{i}(k) \\&+\frac{1}{2} \sum_{j \in[N]} h_i(\gamma) \cdot \delta a_{j}(k)  \left(R_{i j}(k) \cdot \delta a_{j}(k)^{T} +2 r_{i j}(k)\right),
    \end{aligned}
\end{equation} 
where 
\begin{equation*}\label{equ:ROWChar}
    \underset{h_i \in H(\gamma)}{h_i (\gamma)} \begin{cases} -10\gamma + 1, & \text { if } -0.25 \leq \gamma < 0\\ -1.3\gamma + 1 , & \text { if } 0.25\geq \gamma \geq 0 \end{cases}.
\end{equation*} 
$l_i$ and $r_{ij}$ represent the derivatives of $g_i$ with respect to the state and input, respectively: $l_i = \frac{\partial g_i}{\partial \bar{s}}$ and $r_{ij} = \frac{\partial g_i}{\partial \bar{u_j}}$. `

With the inclusion of the behavioural factor vector $H(\gamma)$, the utility of each traffic participant is linked to traffic rules through the behavioural parameter (right-of-way). For instance, when a traffic participant possesses a relatively high right-of-way (e.g., $\gamma > 0$), the corresponding utility for deviations in states from the current states increases, while the utility for deviations in control (action) inputs from the current actions decreases. This encourages the traffic participant to be more assertive. Conversely, the opposite holds true for a lower right-of-way value. The parameters of $h_i (\gamma)$ can be adjusted to represent various traffic-rule-adherence characteristics in the game formulation.

\subsection{Rule-based game solving}
The game is formulated using the linearized dynamics for each traffic participant, considering deviations in state and action space. The utilities of traffic participants are expressed in both linear and quadratic formats. Essentially, the game is transformed into a linearized optimization problem, and the optimal solution of this optimization problem corresponds to the Nash equilibrium of the game \cite{fridovich2020efficient}.

According to \cite{bacsar1998dynamic}, The optima of the optimization problem follows a feedback format:
\begin{equation}\label{equ:feedback}
    {u}_{i}^{*}(k)=-P_{i}^{*}(k) \delta s(k)-\delta \alpha_{i}^{*}(k) + {u}_{i}(k),
\end{equation}
where ${u}_{i}(k)$ is the current actions of ego vehicle and ${u}_{i}^{*}(k)$ is the optimal action at timestep $k$. $P_{i}^{}(k)$ represents the proportional matrix for the feedback solution, and $\alpha_{i}^{}(k)$ denotes the constant offset for the optimal solution. These matrix and constant offset values can be recursively computed using the parameters of the game setup:
\begin{equation}\label{equ:equ_recc1}
\begin{split}
    \left(j_i(\gamma)R^{o}_{i i}+B^{o\, T}_{i} Z^{o+1}_{i} B^{o}_{i}\right) P^{o}_{i}+B^{o\, T}_{i} Z^{o+1}_{i} \sum_{j \neq i} B^{o}_{j} P^{o}_{j}
    \\ =B^{o\, T}_{i} Z^{o+1}_{i} A^{o},
\end{split}
\end{equation}
\begin{equation}\label{equ:equ_recc2}
\begin{split}
    \left(j_i(\gamma)R^{o}_{i i}+B^{o\, T}_{i} Z^{o+1}_{i} B^{o}_{i}\right) \alpha^{o}_{i}+B^{o\, T}_{i} Z^{o+1}_{i} \sum_{j \neq i} B^{o}_{j} \alpha^{o}_{j} 
    \\=B^{o\, T}_{i} \zeta^{o+1}_{i}.
\end{split}
\end{equation}
where $o$ is the iteration index, and $i$ and $j$ indicate the ego vehicle and the surrounding traffic participants. The two mid-term matrices $Z$ and $\zeta$ in (\ref{equ:equ_recc1}) are calculated recursively given $\zeta^{O+1}_{i}=0$ and $Z^{O+1}_{i}=Q^{O+1}_{i}$ as:
\begin{equation}\label{equ:last_cal}
    \left\{
    \begin{aligned}
        Z^{o}_{i} &= F^{o\, T} Z^{o+1}_{i} F^{o} + \sum_{j \in \mathbf{N}} j_i(\gamma)P^{o\, T}_{j} R^{o}_{i j} P^{o}_{j} + j_i(\gamma) Q^{o}_{i}, \\
        \zeta^{o}_{i} &= F^{o\, T} \left(\zeta^{o+1}_{i} + Z^{o+1}_{i} \beta^{o}\right) + \sum_{j \in \mathbf{N}} j_i(\gamma){P^{o\, T}_{j}} R^{o}_{i j} \alpha^{o}_{j},
    \end{aligned}
    \right.
\end{equation}
where
\begin{equation*}%\label{equ:last_cal}
    \left\{
    \begin{aligned}
        F^{o} \triangleq A^{o}-\sum_{i \in \mathbf{N}} B^{o}_{i} P^{o}_{i}, \\
        \beta^{o} \triangleq c^{o}-\sum_{j \in \mathbf{N}} B^{o\, T}_{j} \alpha^{o}_{j}.
    \end{aligned}
    \right.
\end{equation*}
By calculating $P^*$ and $\alpha_i^*$ in an inverse iterative way using (\ref{equ:equ_recc1}) - (\ref{equ:last_cal}), the optimal solution (Nash equilibrium) of the problem can be found \cite{shu2023human}.

The system employs Taylor expansion for linearization and quadratization. As the ego vehicle's future candidate path and speed profile (derived from the ego vehicle's policy) are required to remain close to the current states, the optimal policy is computed iteratively from an initial policy, gradually moving towards the optimum. Hence, to reduce the computational burden, the initial policy could be generated according to traffic rules, rather than the inactive policy.

To integrate this rule-based initial policy setup, the behavioural parameters are utilized to influence the accelerations of the initial policies of vehicles. Specifically, the vehicles' acceleration for the first five timesteps is set to the same value as the behavioural parameters. For example, if a vehicle has a relatively low right-of-way and a behavioural parameter of -0.05, its initial policy's acceleration for the first timestep is -0.05 $m/s^2$. It's important to note that if the future speed generated from the initial policy is less than 0 $m/s$ or exceeds the speed limit, the initial policy becomes an inaction policy.

For pedestrians or cyclists, if their behavioural parameter is 0, the initial policy is set to an inaction policy. If they have an absolute high right-of-way (i.e., their behavioural parameter is 1), the acceleration of the initial policy for the first five timesteps is set to 0.01 $m/s^2$. Therefore, the initial policy is capable of incorporating the precedence of traffic participants according to traffic rules. In most cases, the initial policy is closer to optimal compared to an inaction initial policy.

\begin{figure}
    \centering
    \includegraphics[width=0.35 \textwidth]{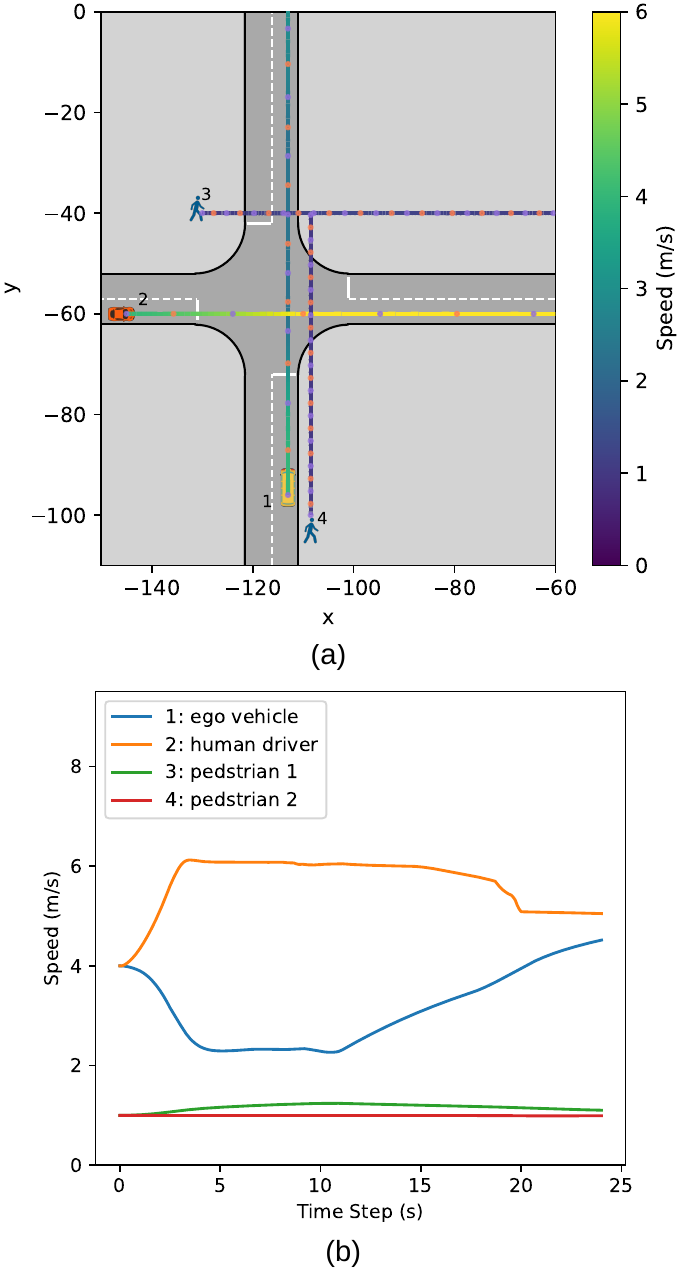}
    \caption{Uncontrolled intersection simulation results}
    \label{Fig: NoRule}
\end{figure}

%\begin{figure}
%    \centering
%    \includegraphics[width=0.35 \textwidth]{Figure/MulHighSpeed.pdf}
%    \caption{Speed profiles of the uncontrolled intersection scenario}
%    \label{Fig: NoRuleSpeed}
%\end{figure}

\section{Simulation and testing result}

The framework is tested using both simulation and real-world vehicle experiments. The simulation scenario is set at an intersection with various traffic rules, including rules leading to definite and indefinite precedence of traffic participants.

Initially, the framework is tested at an uncontrolled intersection where each vehicle has the same right-of-way, and pedestrians and cyclists have absolute high right-of-way. The simulation results are presented in Fig.~\ref{Fig: NoRule}, where sub-figure Fig.~\ref{Fig: NoRule}~(a) is the bird eye view of the simulation result, and Fig.~\ref{Fig: NoRule}~(b) presents the corresponding speed profiles of each traffic participant.
 In Fig.~\ref{Fig: NoRule}~(a), the yellow vehicle and  orange vehicle represent the ego vehicle and human-driven vehicles accordingly, and the blue icon indicates pedestrians. The colour along the trajectories of the traffic participants indicates their speed, with the relationship between traveling speed and colour depicted on the right-side colour bar. Yellow indicates higher speed, while blue indicates slower speed.  The dots along the trajectories indicate the position of the traffic participant at different timesteps, helping to present the relative positions of different traffic participants.

\begin{figure}
    \centering
    \includegraphics[width=0.5 \textwidth]{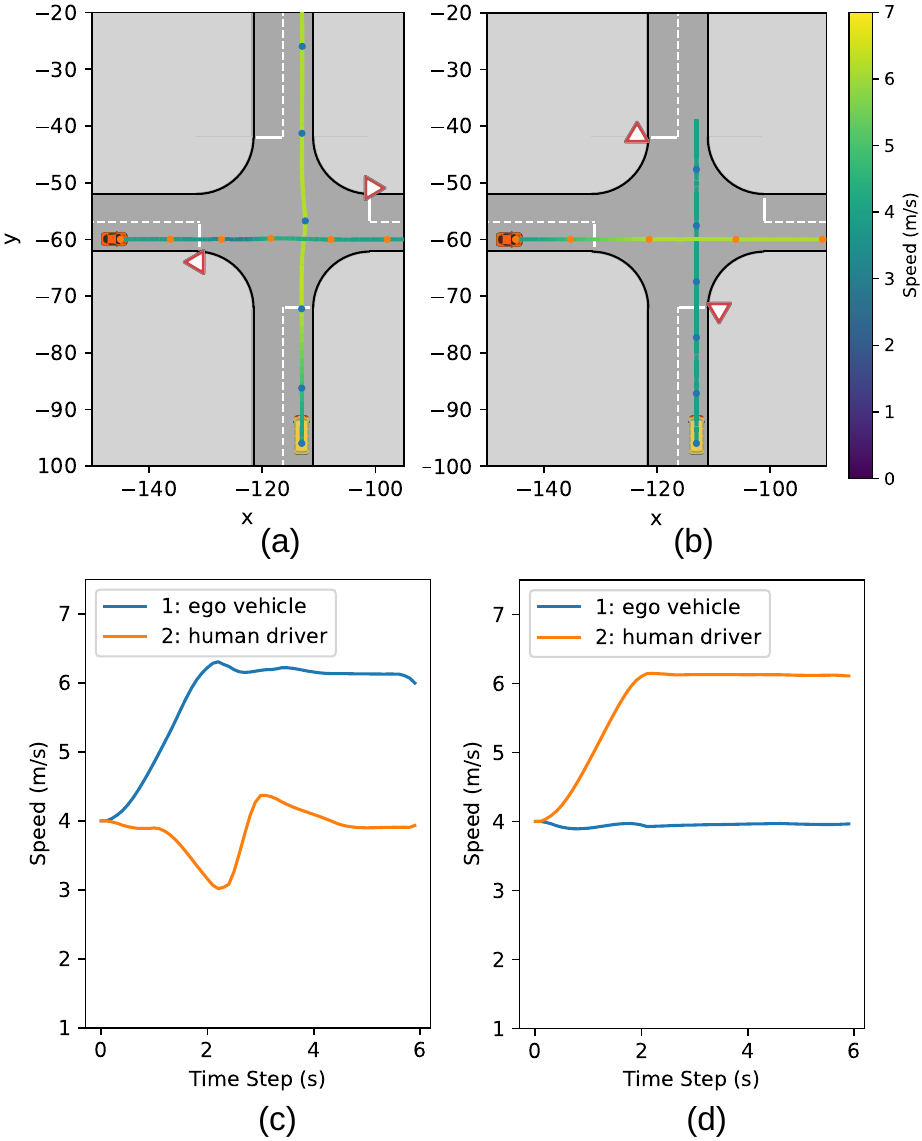}
    \caption{Intersection with yield signs' simulation result}
    \label{Fig: softRule}
\end{figure}

It's important to note that the pedestrian 1 (No.~4 traffic participant) has been opted out by the framework's actor filter since it has a small chance of interaction and is therefore not considered by the framework.

The corresponding speed profiles of this scenario are presented in Fig.~\ref{Fig: NoRule}~(a). The blue, orange, green and red curves correspond to the speed profiles of the ego vehicle (shuttle bus), human-driven vehicle, pedestrian 1 and pedestrian 2. From Fig.~\ref{Fig: NoRule}, it could be observed that, the ego vehicle is able to adjust its speed to properly interact and keep a safe distance with all the traffic participants. 

%It's important to note that the colour of the trajectory of the pedestrian on the right is grey. This indicates that the framework's actor filter has opted out the traffic participant that has a small chance of interaction and is therefore not considered by the framework. The dots along the trajectories indicate the position of the traffic participant at different timesteps, helping to present the relative positions of different traffic participants.

%The corresponding speed profiles of this scenario are presented in Fig.~\ref{Fig: NoRule}~(a). The blue, orange, green and red curves correspond to the speed profiles of the ego vehicle (shuttle bus), human-driven vehicle, pedestrian 1 and pedestrian 2. From Fig.~\ref{Fig: NoRule}, it could be observed that, the ego vehicle is able to first filter out the more critical traffic participant, then be able to adjust its speed to properly interact and keep a safe distance with all the traffic participants. 

\begin{figure}
    \centering
    \includegraphics[width=0.5 \textwidth]{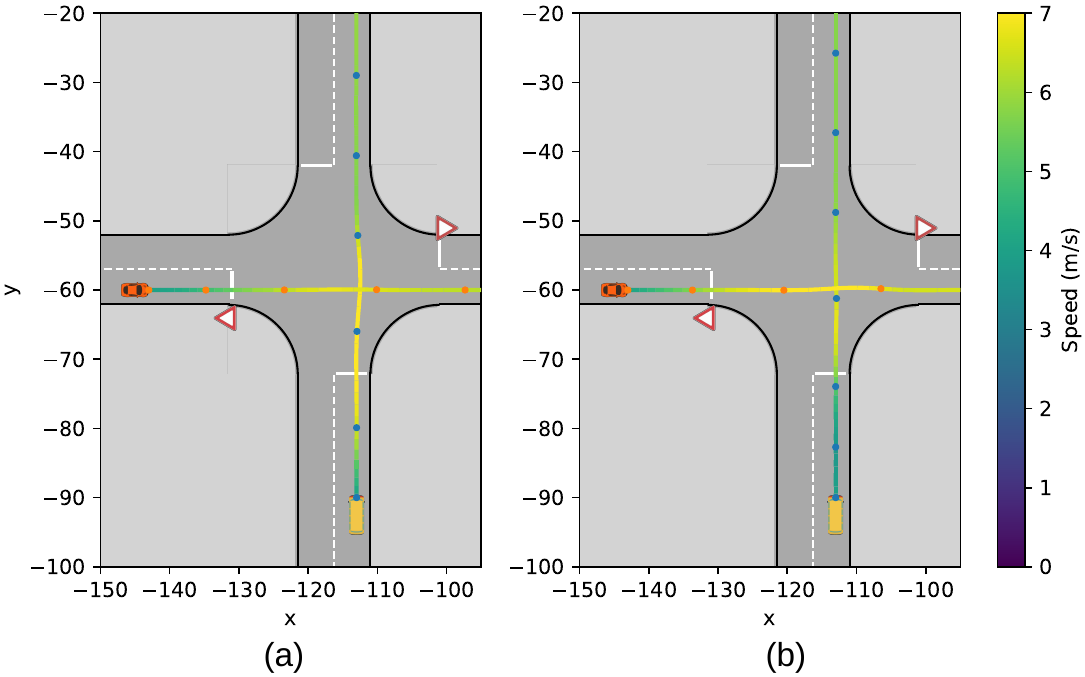}
    \caption{Comparison of traffic rule compliance between proposed and baseline methods at a yield-sign intersection}
    \label{Fig: CompareRule}
\end{figure}

\begin{figure}
    \centering
    \includegraphics[width=0.35 \textwidth]{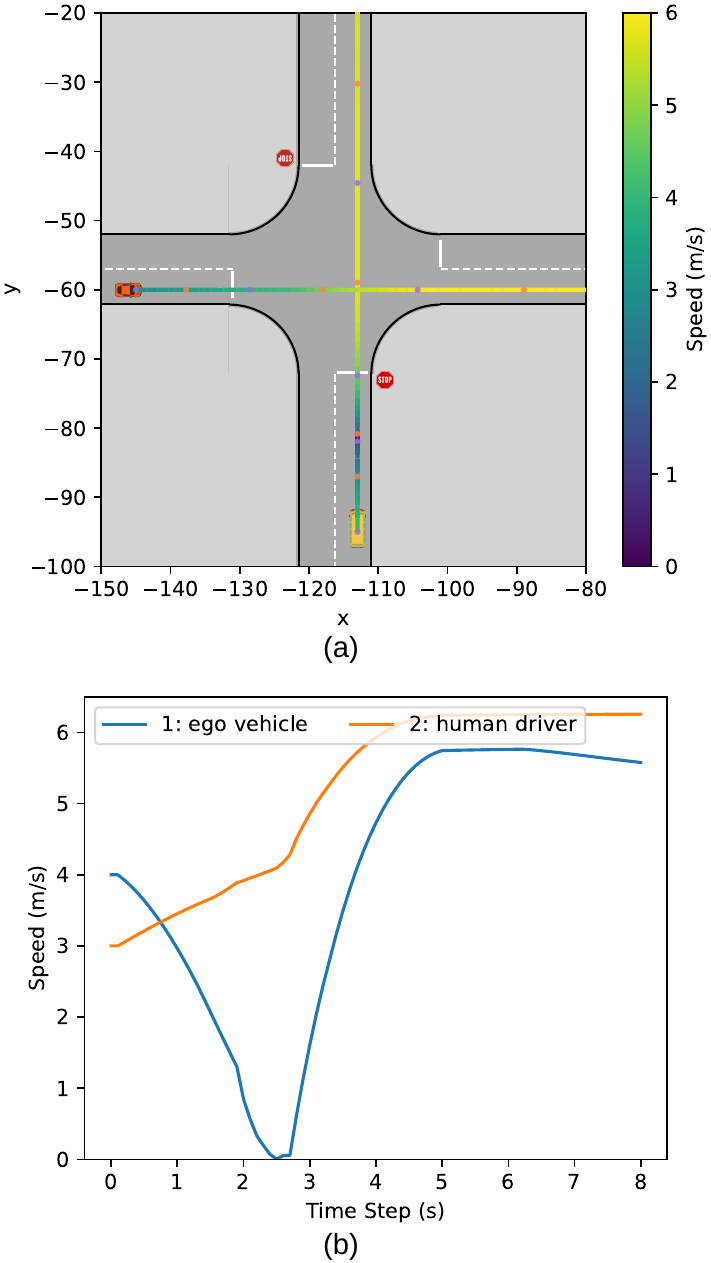}
    \caption{Two-way stop-sign intersection simulation result}
    \label{Fig: StopSignSce}
\end{figure}

The framework's ability to manage more complicated traffic rule scenarios is tested under a 2-way yield sign intersection scenario, as depicted in Fig.~\ref{Fig: softRule}. In sub-figures Fig.\ref{Fig: softRule}(a) and Fig.~\ref{Fig: softRule}~(b), the yellow and orange vehicles represent the ego vehicle and the human-driven vehicle, respectively. The rest of the graphical representations, including lines and dots, are consistent with those shown in Fig.\ref{Fig: NoRule}. Fig.~\ref{Fig: softRule}(c) and Fig.~\ref{Fig: softRule}~(d) display the speed profiles of the traffic participants in scenarios presented in Fig.~\ref{Fig: softRule}(a) and Fig.~\ref{Fig: softRule}~(b), respectively, where the blue and orange lines represent the speed profiles of the ego vehicle and the human-driven vehicle.

In the 2-way yield sign intersection scenario, the right-of-way is clear, but the precedence of the traffic participants is indefinite. In the simulation scenarios depicted in Fig.~\ref{Fig: softRule} (a) and (b), the initial speeds and positions of the traffic participants are the same. However, due to various traffic rules, where the ego vehicle has different right-of-way in those scenarios, its behaviours change accordingly. When the ego vehicle is at the side of the intersection with the yield sign, it maintains a low speed to yield to passing traffic. Conversely, when the other side of the road has the yield sign, the ego vehicle drives with more confidence and passes the intersection first.

The simulation results show that the framework allows the vehicle to incorporate complicated traffic rules by making sound decisions regarding the precedence of traffic participants, while maintaining a safe distance from all other traffic participants.

The proposed framework is compared with a general game-theory-based decision-making framework \cite{shu2023human}, which, while considering strong interactions among traffic participants, are not able to account for traffic rules in urban scenarios. Both frameworks were implemented at a yield-sign intersection in simulation, with the results shown in Fig.~\ref{Fig: CompareRule}.

Fig.~\ref{Fig: CompareRule} (a) presents the decision-making results of the proposed framework, while Fig.~\ref{Fig: CompareRule} (b) depicts the results from the baseline method. These graphical representations maintain consistency with those previously shown in Fig.~\ref{Fig: NoRule}. The images illustrate that the proposed framework successfully integrates the yield-sign-related traffic rules into its decision-making process, allowing the yellow ego vehicle to navigate the intersection first. Conversely, the baseline method does not integrate traffic rule adherence, leading the ego vehicle to yield to another traffic participant, even when it has a relatively higher right-of-way.

The framework's ability to manage strict traffic rules was also tested under a 2-way stop sign intersection. The traffic scenario and the corresponding speed profile are presented in Fig.~\ref{Fig: StopSignSce}~(a) and  Fig.~\ref{Fig: StopSignSce}~(b) respectively. All graphical representations are consistent with those presented in Fig.\ref{Fig: NoRule}.

From the simulation result, it could be observed that the ego vehicle, which is on the side of the intersection with stop signs, is able to follow the traffic rule and stop before the stop sign, then pull into the intersection. 

\begin{figure}
    \centering
    \includegraphics[width=0.38 \textwidth]{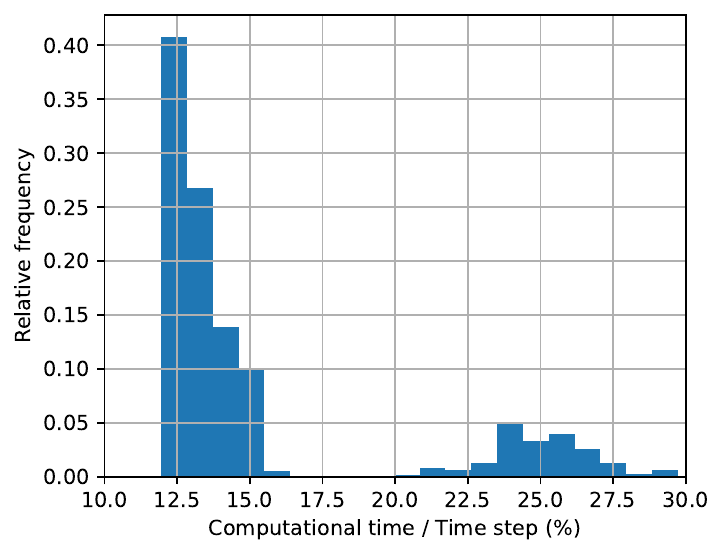}
    \caption{Real-time performance evaluation}
    \label{Fig: Real_time}
\end{figure}

\begin{figure}
    \centering
    \includegraphics[width=0.35 \textwidth]{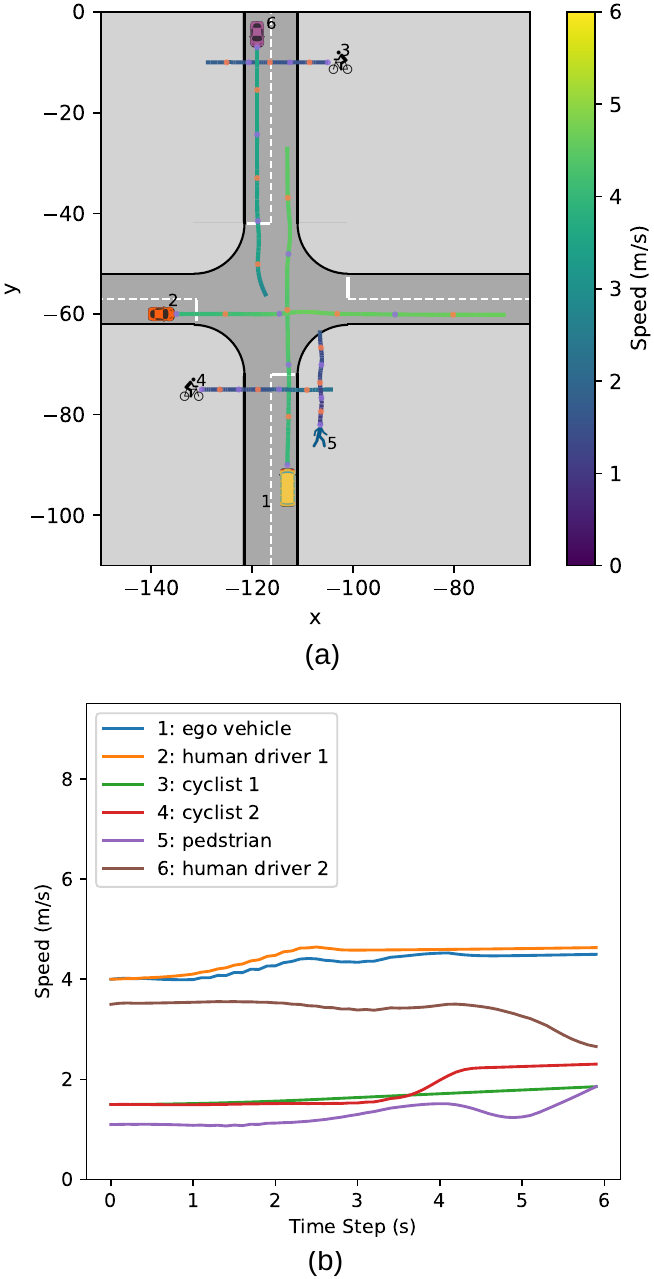}
    \caption{Busy intersection simulation result}
    \label{Fig: Busy_inter}
\end{figure}

%\begin{figure}
%    \centering
%    \includegraphics[width=0.35 \textwidth]{Figure/StopSignSpeed_.pdf}
%    \caption{Speed profiles of intersection with four-way stops}
%    \label{Fig: StopSignSpeed}
%\end{figure}

\begin{figure}
    \centering
    \includegraphics[width=0.38 \textwidth]{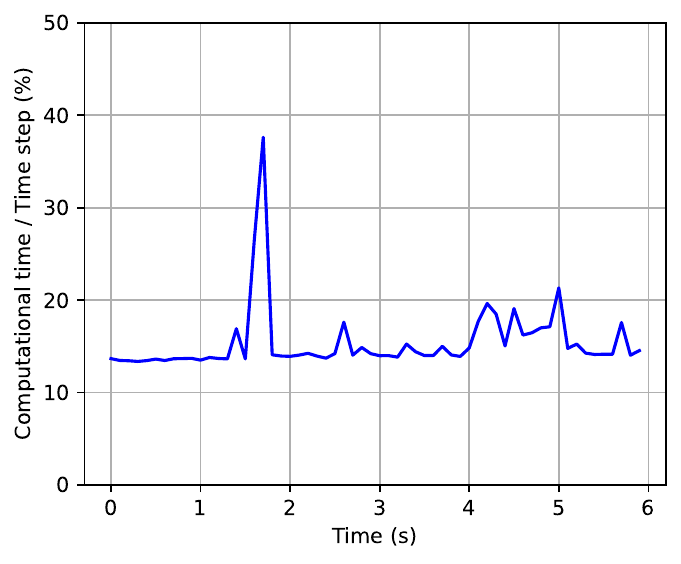}
    \caption{Real-time performance in a busy intersection}
    \label{Fig: Busy_inter_real_time}
\end{figure}

\begin{figure}
    \centering
    \includegraphics[width=0.4 \textwidth]{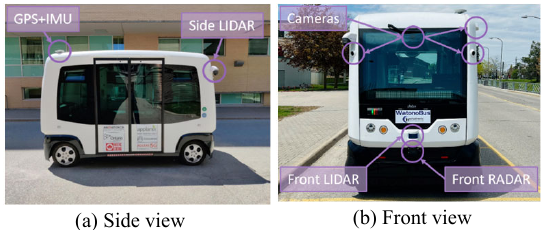}
    \caption{Real-world shuttle bus experiment platform \cite{bhatt2023watonobus}}
    \label{Fig: ShuttleBusLayout}
\end{figure}

One critical factor that ensures safe and efficient decision-making for autonomous vehicles driving in urban scenarios is the real-time performance of the framework. The algorithm was simulated multiple times in various complicated scenarios, running at a frequency of 10 Hz. The time used for generating a solution at each timestep was recorded. Subsequently, the ratio of computational time used to generate a solution relative to each timestep was calculated and documented. Finally, the distribution of the percentage of time used for computation at each timestep is illustrated in Fig.~\ref{Fig: Real_time}. The x-axis represents the percentage of time used for computation for each timestep, while the y-axis represents the relative frequency.

From Fig.~\ref{Fig: Real_time}, it is evident that the framework takes less than 30\% of the time in each timestep for finding a sound path and speed profile. This demonstrates that the algorithm is computationally efficient.

To further evaluate the real-time performance of the proposed decision-making framework, the algorithm was tested at a very busy, unsignalized intersection. The simulation results are displayed in Fig.~\ref{Fig: Busy_inter}, formatted and laid out similarly to those presented in previous simulation scenarios. Fig.~\ref{Fig: Busy_inter} (a) illustrates the paths and corresponding states of all traffic participants from a bird's-eye view, while Fig.~\ref{Fig: Busy_inter} (b) provides a more detailed presentation of the speeds of all traffic participants. The results in Fig.~\ref{Fig: Busy_inter} demonstrate that the ego vehicle is able to generate appropriate paths and corresponding speeds to maintain a safe distance from all human traffic participants while adhering to traffic rules.

\begin{figure*}
    \centering
    \includegraphics[width=0.75 \textwidth]{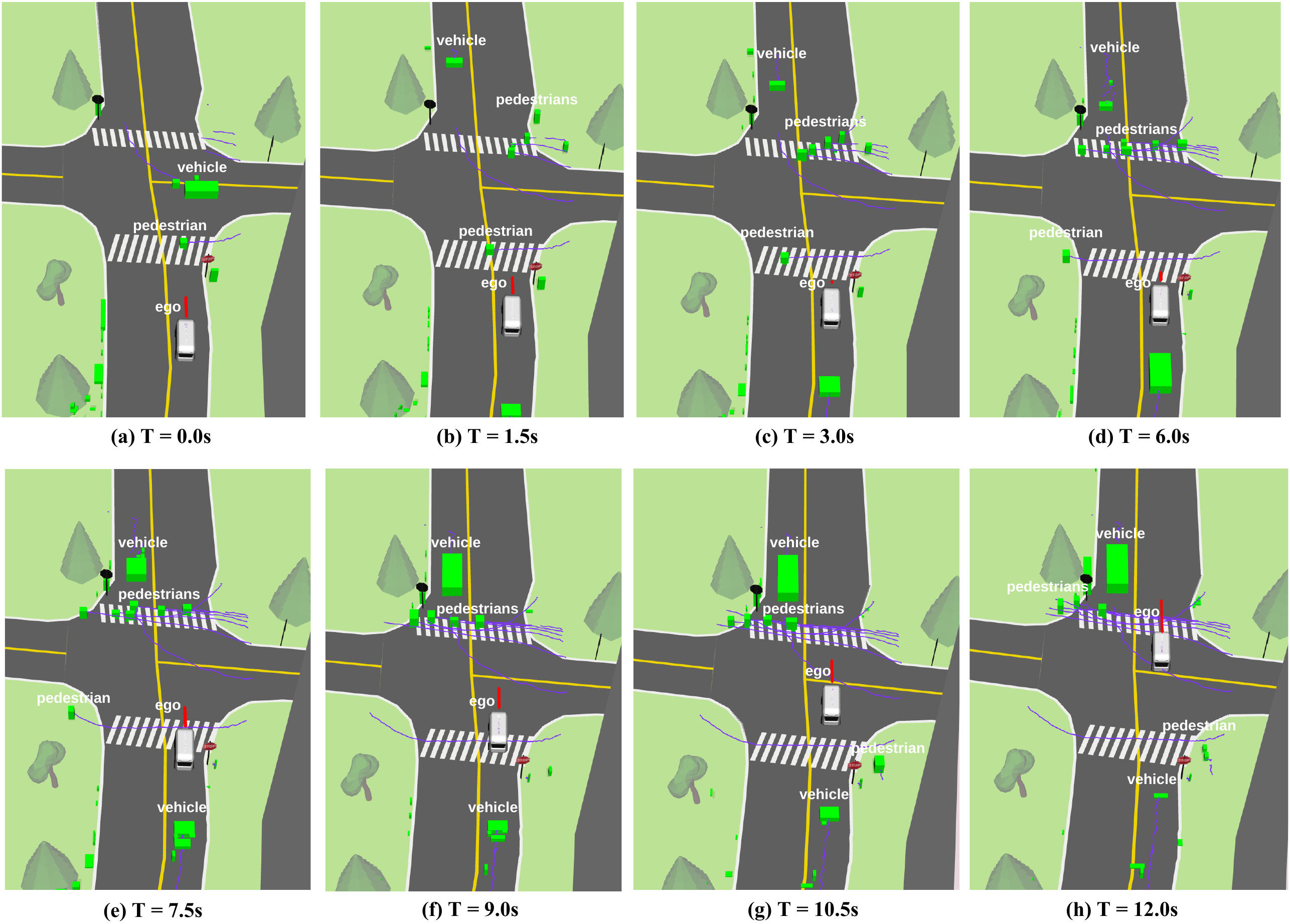}
    \caption{Testing results on the real-world vehicle platform: (a)-(b) ego vehicle approaching the intersection; (c) ego vehicle stopping at stop sign; (d)-(h) ego vehicle initiating movement through the intersection}
    \label{Fig: RealWorldResult}
\end{figure*}

\begin{figure}
    \centering
    \includegraphics[width=0.38 \textwidth]{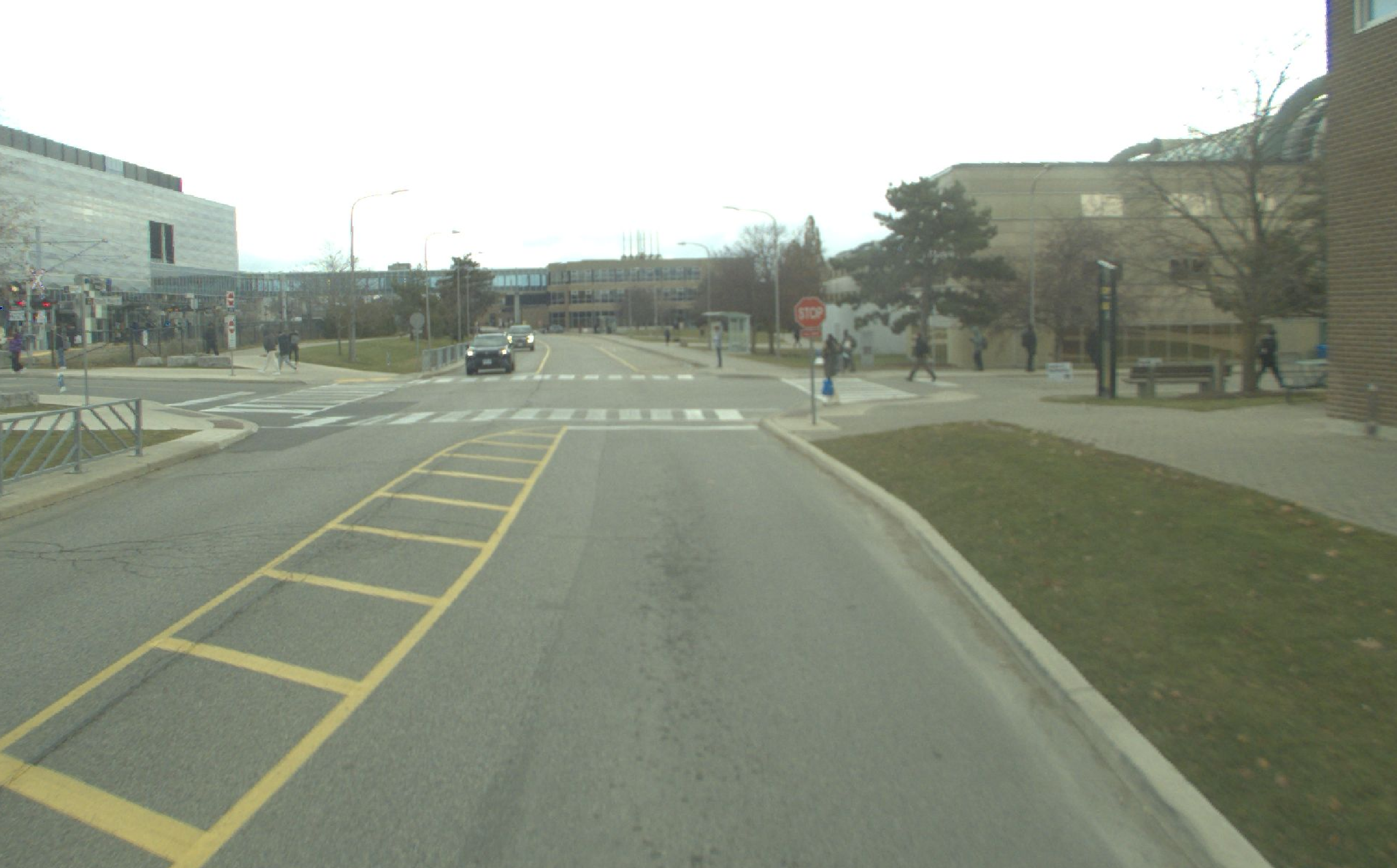}
    \caption{Shuttle bus's front camera view in real-world scenario}
    \label{Fig: CameraShot}
\end{figure}

In the complex and busy intersection simulation scenario, the real-time performance of the proposed framework is demonstrated in Fig.~\ref{Fig: Busy_inter_real_time}. The time used for decision-making at each timestep in the scenario in Fig.~\ref{Fig: Busy_inter} was recorded. Additionally, the proportion of time dedicated to decision-making at each timestep was calculated and plotted over the duration of the simulation. Fig.~\ref{Fig: Busy_inter_real_time} shows that the proposed framework consistently requires less than 20\% of the computation time per timestep and maintains real-time operation throughout the simulation, even under very complicated scenarios.

Finally, the framework is tested on a full-size vehicle (university shuttle bus) platform, as presented in Fig.~\ref{Fig: ShuttleBusLayout}. The vehicle's self-driving system operates under a hierarchical format comprising a perception layer, decision-making layer, and control layer.

The perception layer incorporates various sensors such as cameras, LiDARs, and an inertial measurement unit (IMU), as depicted in Fig.~\ref{Fig: ShuttleBusLayout}. All sensor information is fused and converted into position and speed information of the surrounding traffic participants. The IMU, along with a pre-recorded high-definition map, provides the states of the ego vehicle. Then, all the perception information is sent to the decision-making layer. This layer samples waypoints along the center line of the roads from the high-definition map. These waypoints are then interpolated to generate the nominal path for each participant. Based on these paths, the game-based decision-making model formulates both the path and the corresponding speed profile for the ego vehicle. These outputs are then transmitted to the control layer, which manages the low-level wheel and steering controls.

It is important to note that in real-world applications of self-driving vehicles, the rapidly changing nature of urban traffic scenarios, along with the potential for abrupt changes in pedestrian behavior, can compromise the accuracy of predictions about the actions of surrounding traffic participants. To address this issue, the self-driving vehicle operates its perception and decision-making algorithms at a high frequency (10 Hz), enabling it to respond to sudden changes of actions in a timely manner.
%It is worth noticing that in the real-world self-driving vehicle application, the fast changing nature of the urban traffic scenarios, %the diversed driving patterns of human drivers 
%as well as the possible abrupt change of actions of pedestrians make the estimation of future actions of surrounding traffic participants might not always be fully accurate. The self-driving vehicle addresses this issue by operate the perception and decision-making algorithm in high frequency (10 Hz). Which allows the vehicle to react to sudden changes in a timely manner.

The real-world testing is conducted at a 2-way stop sign intersection, as presented in Fig.~\ref{Fig: RealWorldResult}. Which is visualized based on \cite{poggenhans2018lanelet2}. The intersection features stop signs and walkways at the upper and lower sides, with the left and right sides being the major roads. To better visualize the testing scenario, Fig.~\ref{Fig: CameraShot} presents a real-world video snapshot captured from the front camera of the shuttle bus as it enters the intersection.

In Fig.~\ref{Fig: RealWorldResult}, the white vehicle icon labeled ``ego" represents the ego vehicle (shuttle bus). Other traffic participants (human-driven vehicles and pedestrians), are also labeled and their positions are indicated with green boxes. The red lines represent the planned path for the ego vehicle, while the purple lines depict the historical trajectories of human traffic participants.

The planning results are depicted in Fig.~\ref{Fig: RealWorldResult}, showcasing the planned path of the ego vehicle and the positions of surrounding traffic participants at different time frames. The time frame is indicated at the bottom of each subfigure.

From Fig.~\ref{Fig: RealWorldResult}, it can be observed that the ego vehicle enters the intersection (T = 0$s$), decreases its speed (T = 1.5$s$), and stops in front of the stop sign (T = 3$s$). After the oncoming vehicle completes the left turn and the pedestrian on the same side of the intersection passes, the ego vehicle starts to increase its speed and moves into the intersection (T = 6$s$). It then maintains a low speed to yield to pedestrians on the other side of the intersection (T = 7.5 - 10.5$s$), before increasing its speed when the road is clear (T = 12$s$).

Throughout the planning process, the ego vehicle is able to operate in real time and maintain a safe distance from all traffic participants. The planning results demonstrate that the decisions made by the ego vehicle comply with traffic rules while interacting with other traffic participants.

\section{Conclusion}

The paper proposes a decision-making framework for autonomous vehicles in urban scenarios, which considers interactions as well as traffic rules. The framework utilizes the right-of-way and behaviour parameters of traffic participants to incorporate traffic rules into the game setup. This enables the ego vehicle to consider the influence of traffic rules on each traffic participant while making decisions. The framework employs iterative linear quadratic differential game to allows the vehicle to consider interaction among traffic participants mathematically, by finding the optima of the problem, the Nash equilibrium could be find, which lead to the optimal decision. 

The framework is tested under simulation environment as well as full-size vehicle platform. The results show that the ego vehicle is able to safety interact with surrounding traffic participants while obeying traffic rules.

\bibliographystyle{IEEEtran}
% argument is your BibTeX string definitions and bibliography database(s)
%\bibliography{IEEEabrv,../bib/paper}
\bibliography{bib.bib}

% Generated by IEEEtran.bst, version: 1.14 (2015/08/26)
\begin{thebibliography}{10}
\providecommand{\url}[1]{#1}
\csname url@samestyle\endcsname
\providecommand{\newblock}{\relax}
\providecommand{\bibinfo}[2]{#2}
\providecommand{\BIBentrySTDinterwordspacing}{\spaceskip=0pt\relax}
\providecommand{\BIBentryALTinterwordstretchfactor}{4}
\providecommand{\BIBentryALTinterwordspacing}{\spaceskip=\fontdimen2\font plus
\BIBentryALTinterwordstretchfactor\fontdimen3\font minus \fontdimen4\font\relax}
\providecommand{\BIBforeignlanguage}[2]{{%
\expandafter\ifx\csname l@#1\endcsname\relax
\typeout{** WARNING: IEEEtran.bst: No hyphenation pattern has been}%
\typeout{** loaded for the language `#1'. Using the pattern for}%
\typeout{** the default language instead.}%
\else
\language=\csname l@#1\endcsname
\fi
#2}}
\providecommand{\BIBdecl}{\relax}
\BIBdecl

\bibitem{shi2023trajectory}
X.~Shi and X.~Li, ``Trajectory planning for an autonomous vehicle with conflicting moving objects along a fixed path--an exact solution method,'' \emph{Transportation research part B: methodological}, vol. 173, pp. 228--246, 2023.

\bibitem{gu2015tunable}
T.~Gu, J.~Atwood, C.~Dong, J.~M. Dolan, and J.-W. Lee, ``Tunable and stable real-time trajectory planning for urban autonomous driving,'' in \emph{2015 IEEE/RSJ international conference on intelligent robots and systems (IROS)}.\hskip 1em plus 0.5em minus 0.4em\relax IEEE, 2015, pp. 250--256.

\bibitem{crosato2022interaction}
L.~Crosato, H.~P. Shum, E.~S. Ho, and C.~Wei, ``Interaction-aware decision-making for automated vehicles using social value orientation,'' \emph{IEEE Transactions on Intelligent Vehicles}, vol.~8, no.~2, pp. 1339--1349, 2022.

\bibitem{wang2021interpretable}
H.~Wang, H.~Gao, S.~Yuan, H.~Zhao, K.~Wang, X.~Wang, K.~Li, and D.~Li, ``Interpretable decision-making for autonomous vehicles at highway on-ramps with latent space reinforcement learning,'' \emph{IEEE Transactions on Vehicular Technology}, vol.~70, no.~9, pp. 8707--8719, 2021.

\bibitem{ding2021epsilon}
W.~Ding, L.~Zhang, J.~Chen, and S.~Shen, ``Epsilon: An efficient planning system for automated vehicles in highly interactive environments,'' \emph{IEEE Transactions on Robotics}, vol.~38, no.~2, pp. 1118--1138, 2021.

\bibitem{yuan2024game}
M.~Yuan and J.~Shan, ``Game-theoretic decision-making for autonomous driving vehicles,'' in \emph{Autonomous Vehicles and Systems}.\hskip 1em plus 0.5em minus 0.4em\relax River Publishers, 2024, pp. 269--300.

\bibitem{hang2020human}
P.~Hang, C.~Lv, Y.~Xing, C.~Huang, and Z.~Hu, ``Human-like decision making for autonomous driving: A noncooperative game theoretic approach,'' \emph{IEEE Transactions on Intelligent Transportation Systems}, vol.~22, no.~4, pp. 2076--2087, 2020.

\bibitem{zanardi2021urban}
A.~Zanardi, E.~Mion, M.~Bruschetta, S.~Bolognani, A.~Censi, and E.~Frazzoli, ``Urban driving games with lexicographic preferences and socially efficient nash equilibria,'' \emph{IEEE Robotics and Automation Letters}, vol.~6, no.~3, pp. 4978--4985, 2021.

\bibitem{vicini2022decision}
M.~Vicini, S.~Albut, E.~Gindullina, and L.~Badia, ``Decision making via game theory for autonomous vehicles in the presence of a moving obstacle,'' in \emph{2022 IEEE International Conference on Communication, Networks and Satellite (COMNETSAT)}.\hskip 1em plus 0.5em minus 0.4em\relax IEEE, 2022, pp. 393--398.

\bibitem{shu2023human}
K.~Shu, R.~V. Mehrizi, S.~Li, M.~Pirani, and A.~Khajepour, ``Human inspired autonomous intersection handling using game theory,'' \emph{IEEE Transactions on Intelligent Transportation Systems}, 2023.

\bibitem{zhang2023game}
M.~Zhang, D.~Chu, Z.~Deng, and C.~Zhao, ``Game theory-based lane change decision-making considering vehicle’s social value orientation,'' SAE Technical Paper, Tech. Rep., 2023.

\bibitem{ji2020estimating}
A.~Ji and D.~Levinson, ``Estimating the social gap with a game theory model of lane changing,'' \emph{IEEE Transactions on Intelligent Transportation Systems}, vol.~22, no.~10, pp. 6320--6329, 2020.

\bibitem{jeong2023probabilistic}
Y.~Jeong, ``Probabilistic game theory and stochastic model predictive control-based decision making and motion planning in uncontrolled intersections for autonomous driving,'' \emph{IEEE Transactions on Vehicular Technology}, 2023.

\bibitem{rahmati2021helping}
Y.~Rahmati, M.~K. Hosseini, and A.~Talebpour, ``Helping automated vehicles with left-turn maneuvers: A game theory-based decision framework for conflicting maneuvers at intersections,'' \emph{IEEE transactions on intelligent transportation systems}, vol.~23, no.~8, pp. 11\,877--11\,890, 2021.

\bibitem{liu2022three}
M.~Liu, Y.~Wan, F.~L. Lewis, S.~Nageshrao, and D.~Filev, ``A three-level game-theoretic decision-making framework for autonomous vehicles,'' \emph{IEEE Transactions on Intelligent Transportation Systems}, vol.~23, no.~11, pp. 20\,298--20\,308, 2022.

\bibitem{fan2022food}
J.~Fan, X.~Yao, L.~Zhou, J.~Wood, and C.~Wang, ``Food-delivery behavior under crowd sourcing mobility services,'' \emph{Journal of traffic and transportation engineering (English edition)}, vol.~9, no.~4, pp. 676--691, 2022.

\bibitem{bjornskau2017zebra}
T.~Bj{\o}rnskau, ``The zebra crossing game--using game theory to explain a discrepancy between road user behaviour and traffic rules,'' \emph{Safety science}, vol.~92, pp. 298--301, 2017.

\bibitem{grognard1999global}
F.~Grognard, R.~Sepulchre, and G.~Bastin, ``Global stabilization of feedforward systems with exponentially unstable jacobian linearization,'' \emph{Systems \& Control Letters}, vol.~37, no.~2, pp. 107--115, 1999.

\bibitem{fridovich2020efficient}
D.~Fridovich-Keil, E.~Ratner, L.~Peters, A.~D. Dragan, and C.~J. Tomlin, ``Efficient iterative linear-quadratic approximations for nonlinear multi-player general-sum differential games,'' in \emph{2020 IEEE international conference on robotics and automation (ICRA)}.\hskip 1em plus 0.5em minus 0.4em\relax IEEE, 2020, pp. 1475--1481.

\bibitem{bacsar1998dynamic}
T.~Ba{\c{s}}ar and G.~J. Olsder, \emph{Dynamic noncooperative game theory}.\hskip 1em plus 0.5em minus 0.4em\relax SIAM, 1998.

\bibitem{bhatt2023watonobus}
N.~P. Bhatt, R.~Zhang, M.~Ning, A.~R. Alghooneh, J.~Sun, P.~Panahandeh, E.~Mohammadbagher, T.~Ecclestone, B.~MacCallum, E.~Hashemi \emph{et~al.}, ``Watonobus: An all weather autonomous shuttle,'' \emph{arXiv preprint arXiv:2312.00938}, 2023.

\bibitem{poggenhans2018lanelet2}
F.~Poggenhans, J.-H. Pauls, J.~Janosovits, S.~Orf, M.~Naumann, F.~Kuhnt, and M.~Mayr, ``Lanelet2: A high-definition map framework for the future of automated driving,'' in \emph{2018 21st international conference on intelligent transportation systems (ITSC)}.\hskip 1em plus 0.5em minus 0.4em\relax IEEE, 2018, pp. 1672--1679.

\end{thebibliography}
%\end{thebibliography}

% biography section
% 
% If you have an EPS/PDF photo (graphicx package needed) extra braces are
% needed around the contents of the optional argument to biography to prevent
% the LaTeX parser from getting confused when it sees the complicated
% \includegraphics command within an optional argument. (You could create
% your own custom macro containing the \includegraphics command to make things
% simpler here.)
%\begin{IEEEbiography}[{\includegraphics[width=1in,height=1.25in,clip,keepaspectratio]{mshell}}]{Michael Shell}
% or if you just want to reserve a space for a photo:

%%\newpage

\vspace{11pt}

\begin{IEEEbiography}[{\includegraphics[width=1in,height=1.25in,clip,keepaspectratio]{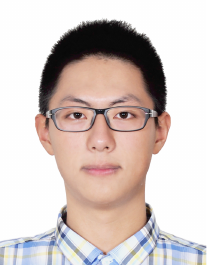}}]{Keqi Shu}
is a PhD candidate in Mechanical and Mechatronics Engineering at the University of Waterloo. His research interests involve interactive planning and decision-making of autonomous vehicles. 
He completed his Master's degree in Mechanical and Mechatronics Engineering in the University of Waterloo, Ontario, Canada, and his B.Sc. degree in Northwestern Polytechnical University, Xi'an Shaanxi, China.
%Before his Masters, he completed his B.Sc. degree in Mechatronics Engineering in Northwestern Polytechnical University, Xi'an Shaanxi, China.
\end{IEEEbiography}

\vspace{-10 mm} 

\begin{IEEEbiography}[{\includegraphics[width=1in,height=1.25in,clip,keepaspectratio]{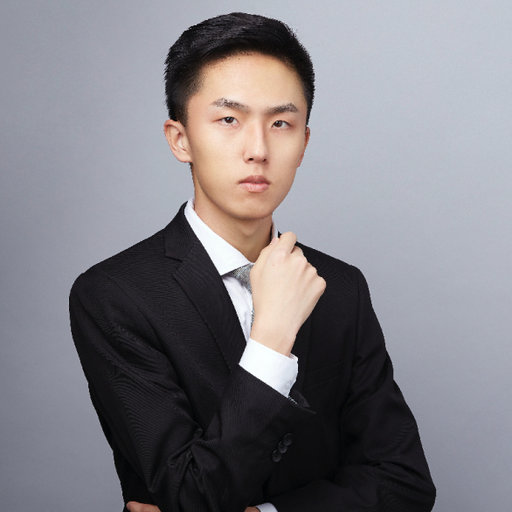}}]{Minghao Ning}
received the B.S. degree in vehicle engineering from the Beijing Institute of Technology, Beijing, China, in 2020. He is currently pursuing the Ph.D. degree with the Department of Mechanical and Mechatronics Engineering, University of Waterloo. His research interests include autonomous driving, LiDAR perception, planning and
control.

\end{IEEEbiography}
\vspace{-10 mm} 

\begin{IEEEbiography}[{\includegraphics[width=1in,height=1.25in,clip,keepaspectratio]{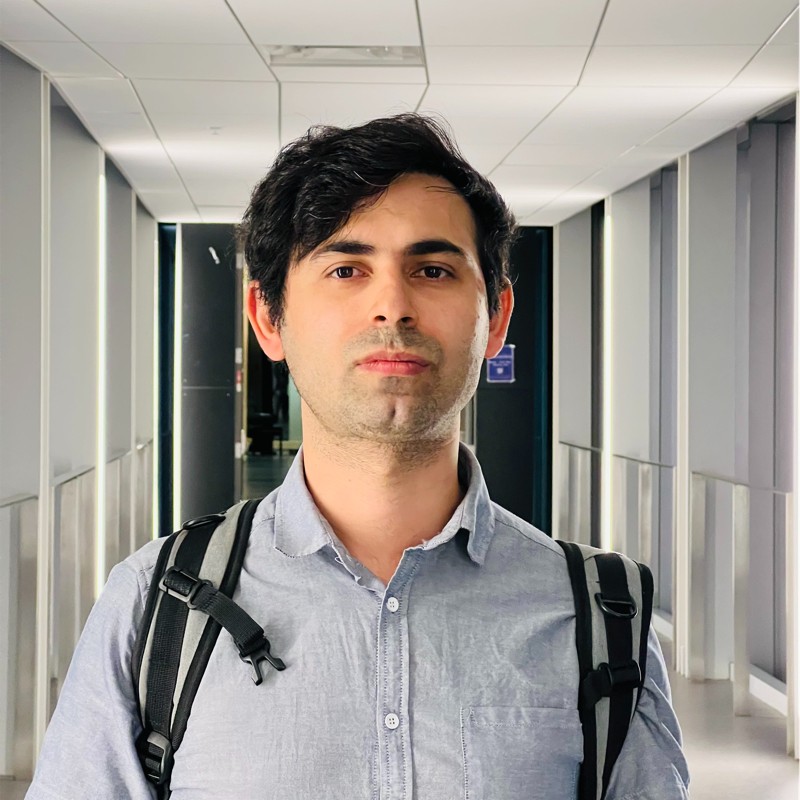}}]{Ahmad Reza Alghooneh}
is a Ph.D. candidate at the University of Waterloo. He received his master’s degree from the University of Tehran, Iran, where he focused on using deep learning to solve perception and object manipulation challenges for humanoid robots. During his master’s, he played a key role in developing the renowned Iranian humanoid robot, SURENA IV. In Canada, Ahmad collaborated closely with General Motors through the Mechatronics Vehicles System laboratory, resulting in a U.S. Patent with the company. His research focuses on sensor fusion, autonomous vehicle perception, and controls. He excels in radar-camera data fusion to address perception challenges in adverse weather conditions.
%Before his Masters, he completed his B.Sc. degree in Mechatronics Engineering in Northwestern Polytechnical University, Xi'an Shaanxi, China.
\end{IEEEbiography}

\vspace{-10 mm}

\begin{IEEEbiography}[{\includegraphics[width=1in,height=1.25in,clip,keepaspectratio]{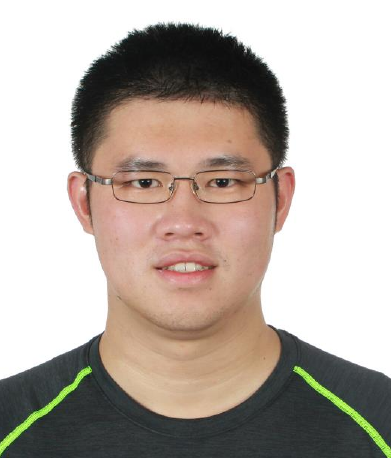}}]{Shen Li}
earned his Ph.D. degree in civil engineering in transportation from the University of Wisconsin–Madison. He is a research associate with the School of Civil Engineering, Tsinghua University, Beijing 100084, China. His research interests include intelligent transportation systems, architecture design of the Connected and Automated Vehicle Highway system, vehicle infrastructure cooperative planning and decision methods, traffic data mining based on cellular data, and traffic operations and management. He is a Member of IEEE.
\end{IEEEbiography}

\vspace{-10 mm} 

\begin{IEEEbiography}[{\includegraphics[width=1in,height=1.25in,clip,keepaspectratio]{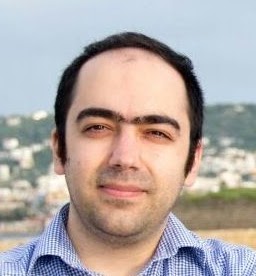}}]{Mohammad Pirani}
is an assistant professor with the Department of Mechanical Engineering, University of Ottawa, Canada. He was a research assistant professor in the Department of Mechanical and Mechatronics Engineering at the University of Waterloo (2022–2023). He held postdoctoral researcher positions at the University of Toronto (2019–2021) and KTH Royal Institute of Technology, Sweden (2018–2019). He received a MASc degree in electrical and computer engineering and a Ph.D. degree in Mechanical and Mechatronics Engineering, both from the University of Waterloo in 2014 and 2017, respectively. His research interests include resilient and fault-tolerant control, networked control systems, and multi-agent systems.
\end{IEEEbiography}

\vspace{-10 mm} 

\begin{IEEEbiography}[{\includegraphics[width=1in,height=1.25in,clip,keepaspectratio]{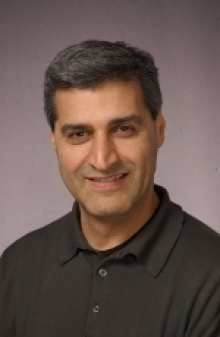}}]{Amir Khajepour}
(Senior Member, IEEE) is a professor of Mechanical and Mechatronics Engineering and the Director of the Mechatronic Vehicle Systems (MVS) Lab at the University of Waterloo. He held the Tier 1 Canada Research Chair in Mechatronic Vehicle Systems from 2008 to 2022 and the Senior NSERC/General Motors Industrial Research Chair in Holistic Vehicle Control from 2017 to 2022. His work has resulted in training of over 150 PhD and MASc students, 30 patents, over 600 research papers, numerous technology transfers, and several start-up companies. He has been recognized with the Engineering Medal from Professional Engineering Ontario and is a fellow of the Engineering Institute of Canada, the American Society of Mechanical Engineering, and the Canadian Society of Mechanical Engineering. 
\end{IEEEbiography}

\vspace{11pt}

%\bf{If you will not include a photo:}\vspace{-33pt}
%\begin{IEEEbiographynophoto}{John Doe}
%Use $\backslash${\tt{begin\{IEEEbiographynophoto\}}} and the author name as the argument followed by %the biography text.
%\end{IEEEbiographynophoto}

\vfill

\end{document}